\documentclass[prl,aps,showpacs,floats,floatfix,superscriptaddress,twocolumn,nofootinbib]{revtex4}

\usepackage{graphicx}
\usepackage{color}
\usepackage{amssymb}
\usepackage{amsmath}

\usepackage{longtable}

\def\laq{\raise 0.4ex\hbox{$<$}\kern -0.8em\lower 0.62ex\hbox{$\sim$}}
\def\gaq{\raise 0.4ex\hbox{$>$}\kern -0.7em\lower 0.62ex\hbox{$\sim$}}

\newlength{\sizeonefig}
\newlength{\sizetwofig}
\newlength{\sizeonefigb}
\newlength{\sizetwofigb}
\begin{document}

\title{A revised upper limit to energy extraction from a Kerr black hole}

\author{Jeremy D. Schnittman} 

\affiliation{Gravitational Astrophysics Laboratory, NASA Goddard Space
  Flight Center, Greenbelt, MD 20771}
\affiliation{Joint Space-Science Institute (JSI), College Park, MD 20742}

\begin{abstract}
We present a new upper limit on the energy that may be extracted
from a Kerr black hole by means of particle collisions in the
ergosphere (i.e., the ``collisional Penrose process''). Earlier work on this
subject has focused largely on particles with critical values of
angular momentum falling into an extremal Kerr black hole
from infinity and colliding just outside the horizon. While
these collisions are able to reach arbitrarily high center-of-mass
energies, it is very difficult for the reaction products to escape
back to infinity, effectively limiting the peak efficiency of such a
process to roughly $130\%$. When we allow one of the initial particles
to have impact parameter $b > 2M$, and thus not get captured by the
horizon, it is able to collide along outgoing trajectories, greatly
increasing the chance that the products can escape. For equal-mass
particles annihilating to photons, we find a greatly
increased peak energy of $E_{\rm out} \approx 6\times E_{\rm in}$. For
Compton scattering, the efficiency
can go even higher, with $E_{\rm out} \approx 14\times E_{\rm in}$,
and for repeated scattering events, photons can both be produced {\it
  and} escape to infinity with Planck-scale energies. 
\end{abstract}


\date{\today}

\maketitle


When two particles scatter or annihilate close to a spinning black
hole, it is possible that one of the products can have negative energy,
necessarily getting captured by the horizon, thereby reducing the black
hole's total mass. This energy is transferred to the other particle,
which then may escape to infinity with more energy than that of the
initial particles. This process, first identified by Penrose in the
context of decay products, has long
been studied as a curious property of Kerr black holes, but is of
questionable astrophysical relevance \cite{Penrose1969}. 

Notably, early work by Wald \cite{Wald1974} showed that a maximum
efficiency of $\approx 121\%$ can be a obtained for a single particle falling
from rest at infinity, then decaying into two equal-mass particles in
the ergosphere, one of which escapes, and one of which is
captured. Piran et al found qualitatively similar results for
a variety of scattering and annihilation processes, with only modest
net energy extraction from the black hole despite arbitrarily large
center-of-mass energies \cite{Piran1975,Piran1977}.

More recently, the subject has received renewed interest following a
paper by Banados, Silk, and West (BSW)~\cite{Banados2009} that proposed to
use extremal Kerr black holes to probe Planck-scale energies
by colliding particles on
critical trajectories just outside the horizon. Numerous subsequent
papers pointed out that, while very large center-of-mass energies are
indeed obtainable, the subsequent redshift of the escaping particles
ensures that the net efficiency of the Penrose process remains limited
to only about $130\%$ \cite{Jacobson2010,Banados2011,Harada2012,Bejger2012,McWilliams2013}.

One feature that all these papers have in common is their analytic
approach to the problem. Based on strong symmetry arguments, they
focus almost exclusively on very specific geodesic trajectories in the
equatorial plane, usually taking the limit as the point of collision
approaches the horizon, where the center-of-mass energy is highest. In
a companion work \cite{Schnittman2014}, we have taken a different
approach, calculating numerically the distribution
function of dark matter particles around a Kerr black hole using a
fully three-dimensional Monte Carlo code to integrate a huge sample of
random geodesic orbits for both particles and annihilation
photons \cite{Schnittman2013}. In doing so, we consistently found a
small number of photons that managed to escape the black hole with
energies well in excess of the theoretical bounds of
\cite{Harada2012,Bejger2012}. 

In this Letter, we focus on these extreme events, again returning to
analytic methods to understand the properties of the colliding
particles, their products, and the dependence on physical parameters
such as the black hole spin. We arrive at the somewhat startling conclusion
that for annihilation between equal-mass particles falling from rest
at infinity, the escaping photons can attain energies equal to {\it six
times} the total rest mass energy of the incoming particles.
For Compton scattering between a photon and massive
particle, the efficiency can reach nearly $1400\%$! Finally, we
show how repeated scattering events can increase the energy of a
photon---and the net efficiency of the Penrose process---almost without
bound.

We begin with a description of geodesic trajectories in the equatorial
plane around a Kerr black hole. From the normalization constraint
$g_{\mu \nu}p^\mu p^\nu=-m^2$ we can write down an effective potential:
\begin{equation}
V_{\rm eff}(r) = k\frac{M}{r}+\frac{\ell^2}{2r^2}
+\frac{1}{2}(-k-\varepsilon^2)\left(1+\frac{a^2}{r^2}\right)
-\frac{M}{r^3}(\ell-a\varepsilon)^2\, ,
\end{equation}
where $r$ is the radius in Boyer-Lindquist coordinates, $\ell$ and
$\varepsilon$ are the particle's specific angular momentum and energy,
$M$ and $a$ are the black hole mass and spin, and $k=0$ for photons and
$k=-1$ for massive particles. We set $G=M=c=1$ throughout this
paper. For a specific choice of $a$, $\ell$, $\varepsilon$, and $k$,
we can solve for the radial turning points by setting 
$V_{\rm eff}(r)=0$. 

\begin{figure}
\includegraphics[width=0.45\textwidth,clip=true]{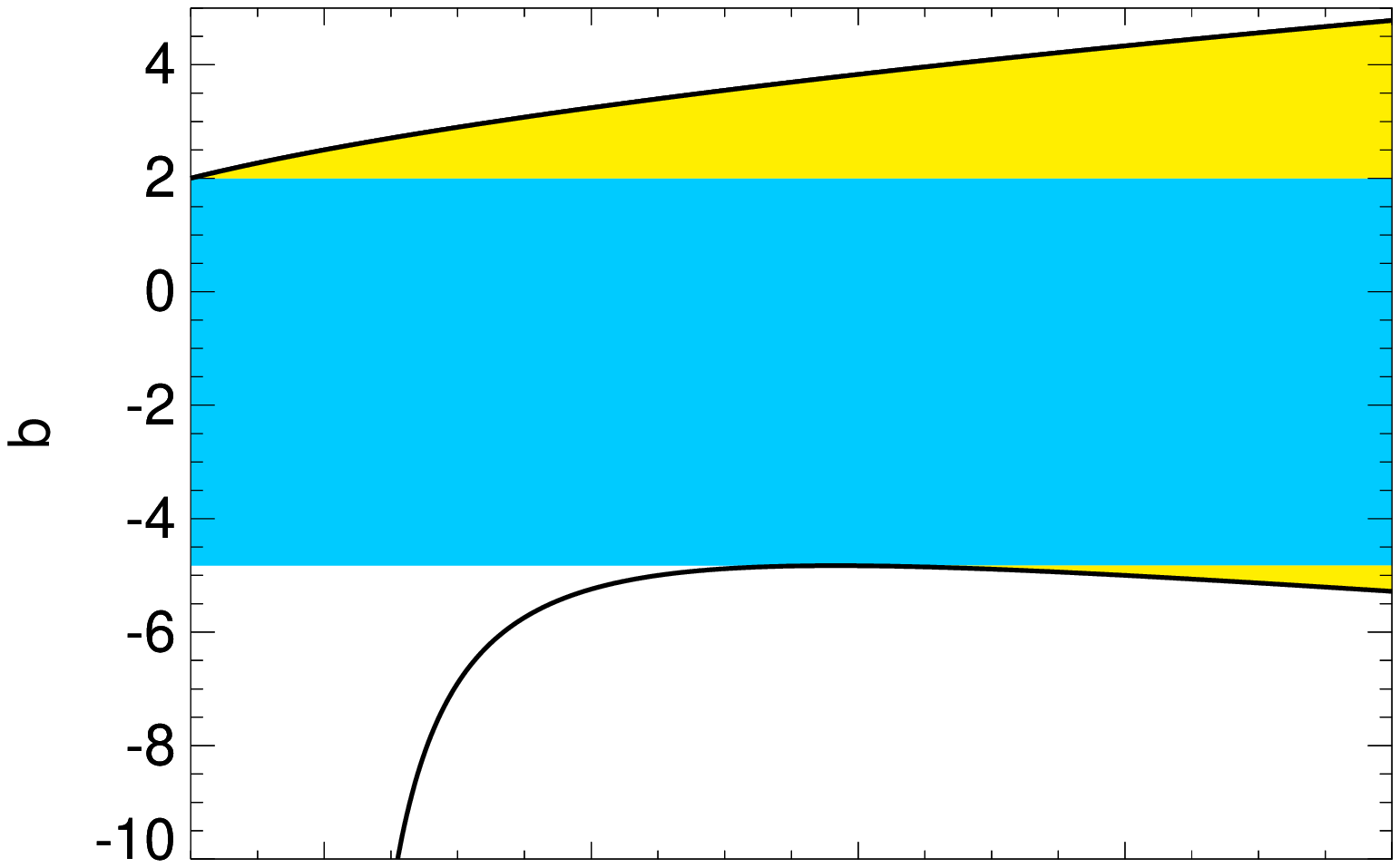}\\
\vspace{-1cm}
\includegraphics[width=0.45\textwidth,clip=true]{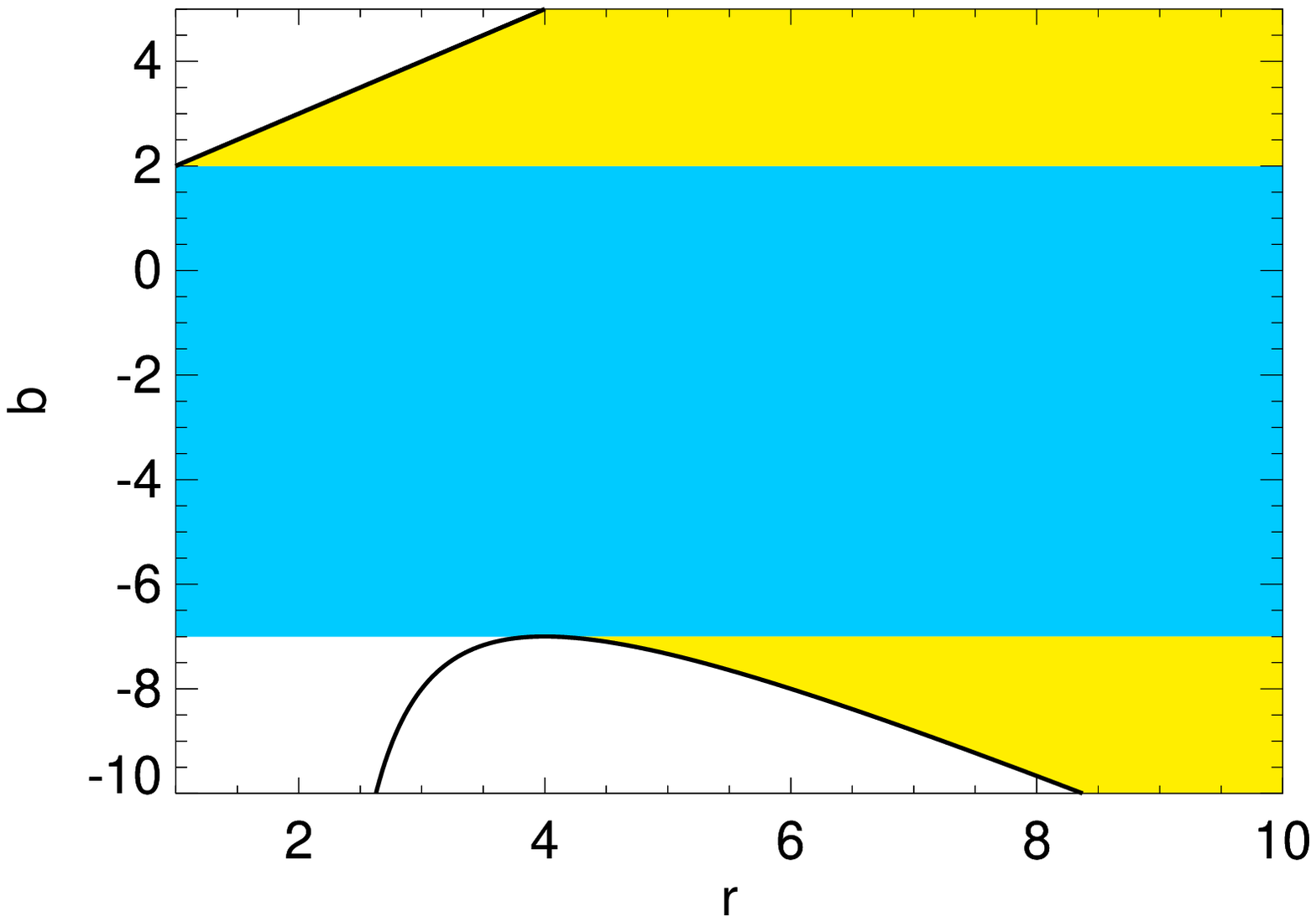}
\caption{\label{fig:Veff} Radial turning points in the effective potential $V_{\rm
    eff}(r,b)$ for ({\it top}) massive and ({\it bottom}) massless
  particles, for a black hole with maximal spin $a=1$. Any particle in
  the yellow region can escape from the black
  hole, but in the blue regions, only particles with outgoing
  radial velocities can escape. The static limit is located at $r=2$ and
  the horizon is at $r=1$.}
\end{figure} 

Figure \ref{fig:Veff} shows these turning points as a function of
the impact parameter $b\equiv \ell/\varepsilon$ for both massless and massive particles, for
maximal spin $a=1$. For the
massive particles we set $\varepsilon=1$, corresponding to a particle
at rest at infinity. One can imagine a massive particle incoming
from the right with $b < -2(1+\sqrt{2})$ or $b > 2$, reflecting
off the centrifugal potential barrier and returning back to infinity (yellow
regions). Alternatively, if the impact parameter is small enough
(i.e., $-2(1+\sqrt{2}) < b < 2$), the particle will get captured by
the black hole. Due to frame-dragging, the cross section for capture
is much greater for incoming particles with negative angular
momentum.

For a given $\ell=p_\phi$ and $\varepsilon=-p_t$, the radial momentum $p_r$ can be
determined from the normalization condition $p_\mu p_\nu g^{\mu
  \nu}=k$:
\begin{equation}
p_r = \pm [g_{rr}(k-g^{tt}\varepsilon^2+2g^{t\phi}\ell\varepsilon
  -g^{\phi \phi}\ell^2)]^{1/2}\, ,
\end{equation}
and the sign of the root is chosen depending on the criterion described
below.

Let us now consider a simple reaction between two particles with
momenta $\mathbf{p}^{(1)}$ and $\mathbf{p}^{(2)}$ that collide to
produce two particles with $\mathbf{p}^{(3)}$ and
$\mathbf{p}^{(4)}$. Constraining all trajectories to the
equatorial plane,
we have to solve for six unknown momentum components. Conservation of
total momentum provides three constraints, and the normalization
conditions on the two daughter particles provide two more. We are
left with a single free parameter: the angle $\psi$ between
$\mathbf{p}^{(3)}$ and the $\hat{\mathbf{r}}$-direction in the
center-of-mass frame.

After solving for the momenta $\mathbf{p}^{(3)}$ and
$\mathbf{p}^{(4)}$ of the products, we determine if they can escape
the horizon based on the effective potential barriers and the sign of
the radial velocity component, as evident in Figure
\ref{fig:Veff}. Scanning over all values of the angle $\psi$, we 
solve for the maximum energy of escaping particles as a function of
radius. Then the peak efficiency of the reaction is defined as
$\eta_{\rm max} = \varepsilon^{(3)}_{\rm max}/
(\varepsilon^{(1)}+\varepsilon^{(2)})$. 
Interestingly, as pointed out by \cite{Piran1977,Bejger2012}, the
highest energy photons that can escape are initially
ingoing with $b_3 >2$, giving $\eta_{\rm max} \approx
1.3$, while those photons with initially outward-directed trajectories
typically have $\eta_{\rm max} < 0.5$.

BSW and much of the subsequent work on this topic
focus predominantly on reactions between equal-mass
particles falling from rest at infinity with critical
values of the impact parameters ($b_1=2$,
$-2(1+\sqrt{2}) \le b_2 \le 2$), which lead
to divergent center-of-mass energies just outside the horizon when
$a=1$. Yet the products of such
reactions are either immediately captured by the black hole, or lose
most of their energy due to gravitational redshift before reaching an
observer \cite{Banados2011,Harada2012,Bejger2012,McWilliams2013}. 

\begin{figure}
\includegraphics[width=0.45\textwidth,clip=true]{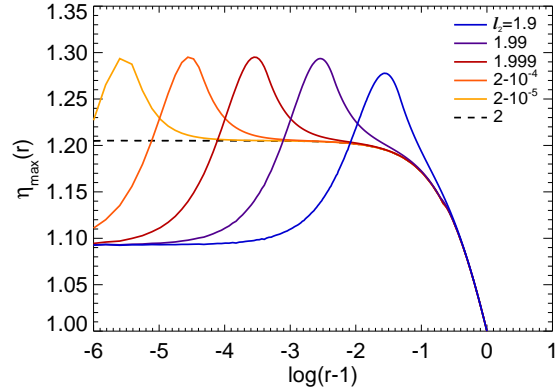}
\caption{\label{fig:Bejger2} Peak efficiency for annihilation of equal-mass particles
  falling from rest at infinity, as a function of the radius at which
  the annihilation occurs. The impact parameter $b_1=2$ is fixed at
  the critical value, and $b_2$ varies. The black hole spin is
  maximal: $a=1$. (Compare to Fig.\ 2 of Ref.\ \cite{Bejger2012})} 
\end{figure} 
In Figure \ref{fig:Bejger2} we plot $\eta_{\rm max}(r)$ for the same
parameters as in Figure 2b of Ref.~\cite{Bejger2012}, and
reassuringly get identical results. In all cases, $m_1=m_2=1$,
$m_3=m_4=0$, $\varepsilon_1=\varepsilon_2=1$, $a=1$,
$b_1=2$, and $p_r^{(1)}, p_r^{(2)}<0$. One interesting result is
that, while the center-of-mass energy increases with smaller $r$, the
peak efficiency is reached somewhat outside the horizon, due to the
greater probability of escape. Note also that the efficiency can only
surpass unity for $r<2$, because the Penrose process only works inside
the ergosphere. The dashed line, for $b_2=b_1=2$, is physically
equivalent to spontaneous breakup of a single particle into two
photons, as in \cite{Wald1974}. 


While exploring the distribution function and
annihilation products of dark matter
around Kerr black holes \cite{Schnittman2014}, we consistently found
escaping photons with energies greater than the limits of Bejger et
al.~\cite{Bejger2012}. An important clue to understanding these seemingly impossible
photons came from Figure \ref{fig:Bejger2}: the greatest efficiency
actually comes from trajectories somewhat outside the
horizon. When considering initial particles with $b > 2$ that never
reach the horizon, it soon became obvious that they can reflect off
the effective potential barrier and escape with $p_r>0$. When
colliding with particles on incoming trajectories, these
``super-critical'' particles contribute some positive radial
momentum to the center-of-mass frame, thereby increasing the range of
angles---as measured in the coordinate frame---accessible to the product
photons. This in turn greatly increasing the chance that high-energy
products can escape to infinity (even if they are initially on ingoing
trajectories).

\begin{figure}
\includegraphics[width=0.45\textwidth,clip=true]{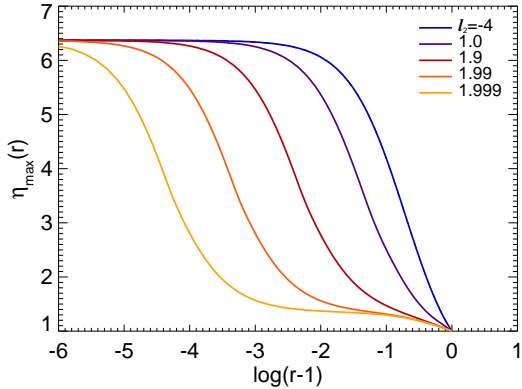}
\caption{\label{fig:peakE_new} Peak efficiency for annihilation of equal-mass particles
  falling from rest at infinity, as a function of the radius at which
  the annihilation occurs. Unlike in Fig.\ \ref{fig:Bejger2}, here we
  allow $p_r^{(1)}>0$, which greatly increases the fraction and energy
  of escaping photons. The impact parameter $b_1=2$ is fixed at
  the critical value, and $b_2$ varies.}
\end{figure} 

In Figure \ref{fig:peakE_new} we again plot the peak efficiency for
annihilation reactions between two equal-mass particles with
$\varepsilon_{1,2}=1$. Yet now we set $b_1$ to be infinitesimally
larger than the critical value, allowing $p_r^{(1)}>0$. The difference
is quite profound. Not only is the peak efficiency much greater than
previously calculated, but it is also attainable for a wider range of
$b_2$ values. 

\begin{figure}
\includegraphics[width=0.45\textwidth,clip=true]{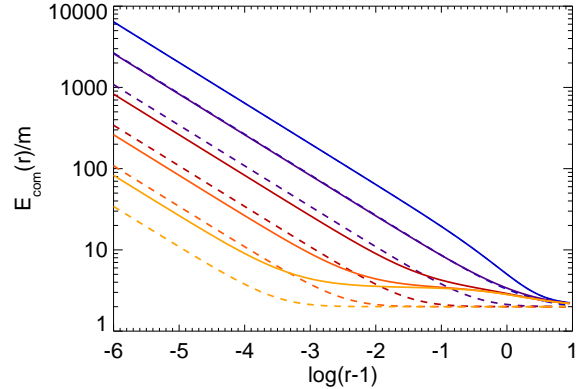}
\caption{\label{fig:Ecom} Center-of-mass energy for the annihilation reactions shown in
  Figs.\ \ref{fig:Bejger2} and \ref{fig:peakE_new}. The color scheme
  is as in Fig.\ \ref{fig:peakE_new}, with solid lines corresponding
  to $p_r^{(1)}>1$ and dashed lines for $p_r^{(1)}<1$. The
  highest COM energies naturally occur for the prograde-retrograde
  collisions, which are closest to ``head-on.''}
\end{figure} 

To better understand the cause of this increase, in
Figure \ref{fig:Ecom} we plot the center-of-mass energy $E_{\rm com}/m$
as a function of radius, for both $p_r^{(1)}>0$
(solid curves) and $p_r^{(1)}<0$ (dashed curves). The values for
$b_2$ and the color scheme are
identical to Figure \ref{fig:peakE_new}. Interestingly, the peak
center-of-mass energy is only a factor of $\sim 2$ higher for outgoing
particles relative to ingoing particles. Thus we conclude that the
huge increase in efficiency in our case is due to a higher escape
probability for the products, rather than a much higher initial photon
energy. 

\begin{figure}
\includegraphics[width=0.45\textwidth,clip=true]{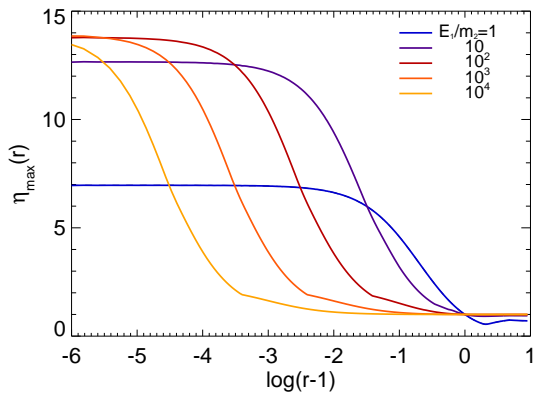}
\caption{\label{fig:peakEC} Peak efficiency for Compton-like scattering events between a
  photon with $b_1=2$, $p_r^{(1)}>1$ and a massive particle
  falling from rest at infinity with rest mass $m_2$ and impact parameter
  $b_2=-2(1+\sqrt{2})$.} 
\end{figure}

We also considered Compton-like scattering between a photon and
massive particle in a similar geometry to the annihilation mechanism,
with $b_1$ just above the critical value,
$p_r^{(1)}>0$, $b_2 = -2(1+\sqrt{2})$, and $p_r^{(2)}<0$. The
results are shown in Figure \ref{fig:peakEC}, where the different
curves correspond to different photon energies $\varepsilon_1$, and the
efficiency is still defined as
$\varepsilon_3/(\varepsilon_1+\varepsilon_2)$\footnote{The ``kinks''
  in the curves correspond to different roots of the quadratic
  equation that arise when solving for the momentum constraints and
  escape conditions for $\mathbf{p}^{(3)}$}. When $\varepsilon_1\gg
m_2$, we find the escaping photons
can reach energies roughly {\it 14 times} that of the incoming
photons. 

As noted above, the highest-energy photons that are
able to escape are actually emitted with {\it ingoing} radial
momentum before reflecting off the potential barrier and then reaching
infinity. This allows for the fortunate possibility of multiple
scattering events with the same photon: each time, the photon first
reflects off the potential barrier so that $p_r^{(1)}>0$, then
scatters off a new infalling particle with $\varepsilon_2=1$,
$b_2=-2(1+\sqrt{2})$, and the resulting photon has
$p_r^{(3)}<0$ and $b_3>2$, allowing the process to
repeat. 

Thus it is possible to reach photon energies on the order of 
$m_2\, 10^N$ after $N$ scattering events, for a net efficiency of
$\eta \approx 10^N/N$. The only limitation, as can be seen in Figure
\ref{fig:peakEC}, is that the radius at which such high-efficiency
reactions ($\eta \gtrsim 10$) can occur moves steadily inward with
each subsequent scattering event. We call this radius 
$r_{{\rm crit},N}$ and find that 
$\log_{10}(r_{{\rm crit}, N}-1)\approx -1-N$. 

\begin{figure}
\includegraphics[width=0.45\textwidth,clip=true]{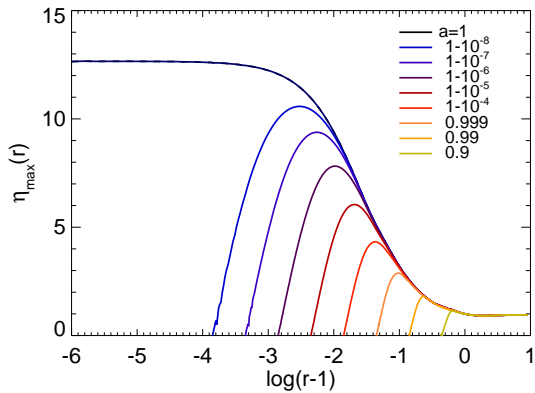}
\caption{\label{fig:peakEC_spin} Spin dependency for efficiency of the scattering events
  plotted in Fig.\ \ref{fig:peakEC} with $\varepsilon_1/m_2=10$.}
\end{figure}

At the same time, each scattering event deposits some negative angular
momentum into the black hole, reducing its spin. This in turn changes
the effective potential, and increases both the impact parameter and
turning radius for the critical trajectories required for
high-efficiency scattering events. In the limit of $\epsilon\equiv
(1-a) \ll 1$, we find the critical value is $b \approx
2(1+\epsilon^{1/2})$, which corresponds to a turning radius of $r_{\rm
  turn}\approx 1+2\epsilon^{1/2}$. Eventually, the spin is reduced to
a point where the turning radius is outside of $r_{\rm crit}$ and the
efficiency begins to suffer (see Fig.\ \ref{fig:peakEC_spin}). 

Following \cite{Thorne1974}, we can write an expression for the spin
evolution due to capture of a particle with angular momentum $\delta
J$ and energy $\delta E$:
\begin{equation}
\delta a = -2a\frac{\delta E}{M}+\frac{\delta J}{M^2}\, .
\end{equation}
Assuming each scattering event has an
efficiency of $\eta=10$, and we start with $a=1$, the change in spin
due to the $N^{th}$ event is to leading order\footnote{While each
  post-scattering particle $m_4$ deposits a greater amount of negative
  angular momentum, it also deposits more negative energy, so the net
  change in spin is relatively constant.}
\begin{equation}
\delta \epsilon_N \approx 2\frac{m_2}{M}\left(
2+\sqrt{2}+\epsilon_N^{1/2}10^{N+1}\right)\, ,
\end{equation}
where $1-\epsilon_N$ is the spin just before the $N^{th}$ event.
Since the condition $r_{\rm turn} < r_{\rm crit}$ requires
$\epsilon_N^{1/2}10^{N+1}<1/2$, we can write
\begin{equation}
\epsilon_{N+1} \approx (4+2\sqrt{2})N\frac{m_2}{M}\, , 
\end{equation}
which in turn gives the maximum number of high-efficiency scattering events:
$N_{\rm max} \approx \log_{10}(M/m_2)^{1/2}$.

Taking $m_2$ to be the electron mass and $M=10M_\odot$ gives $N_{\rm
  max} \approx 30$, for a peak energy of $10^{26}$ GeV. Thus
photons undergoing repeated Compton scattering events could not only
far surpass the Planck energy scale, but these hyper-energetic photons
could even escape to an observer at infinity! 

By simply relaxing a single assumption about the inital trajectories
of colliding particles, we have identified a mechanism that allows
high-energy particles to be produced by the collisional Penrose
process and also escape to infinity. It is entirely possible that
further exploration of parameter space will find even more extreme
cases that have not yet been considered in this study.

At the same time, it is eminently clear that the most extreme effects
will have little or no application to astrophysical processes, where
random particle trajectories and 
realistic black hole spins will likely never reach the critical levels
required for such events. For
example, quantum effects and Hawking radiation will
naturally spin down even an isolated extremal black hole
\cite{Page1976}, and gravitational radiation from the scattering
particles could further degrade the
spin \cite{Berti2009} (although a carefully constructed steady-state
stream of incoming particles might avoid this problem by removing
the time-varying quadrupole moment). A more astrophysically realistic
calculation of potentially observable signatures from dark matter
annihilations is given in \cite{Schnittman2014}.

One might imagine an advanced civilization that can create extremal
black holes and harness them as energy storage devices or
particle accelerators. Of course, there is the problem of
spinning down the black hole in the process, so that only a single extreme event
might ever be achieved \cite{Berti2009}, but this too could be
avoided by providing a simultaneous source of positive angular momentum via
accretion. The net efficiency would suffer, but such is
the cost of probing Planck-scale physics in the laboratory. 

It is a pleasure to acknowledge helpful discussions with A.\ Buonanno,
T.\ Jacobson, and J.\ Silk. Special gratitude is due to
HKB''H for providing us a world full of elegant wonder and beauty. 
This work was partially supported by NASA grants ATP12-0139 and
ATP13-0077. 

\renewcommand{\prd}{\emph{Phys. Rev. D}}

\end{document}